\documentclass[graybox]{svmult}
\usepackage{mathptmx}       % selects Times Roman as basic font

\usepackage{graphicx}        % standard LaTeX graphics tool
\usepackage{epstopdf}
\usepackage{pdfsync}
\usepackage{multicol}        % used for the two-column index
\usepackage[bottom]{footmisc}% places footnotes at page bottom
             \font\sevenrm=cmr7

          \font\sixrm=cmr6       

\def\teq#1{$\, #1\,$}                         % text equation
\def\machson{{M}_{\hbox{\sixrm S}}}

\def\thetascatt{\theta_{\hbox{\sevenrm scatt}}}
\def\thetaBone{\Theta_{\hbox{\sevenrm Bf1}}}
\def\thetaBnone{\Theta_{\hbox{\sevenrm Bs1}}}

\def\betaHTone{\beta_{\hbox{\sevenrm 1HT}}}
\def\uHTone{u_{\hbox{\sevenrm 1HT}}}
\def\Rlc{R_{\hbox{\sevenrm lc}}}
\def\lambar{\lambda\llap {--}_{\rm c}} 
\def\emax{\varepsilon_{\hbox{\sixrm MAX}}} 
\def\epsinj{\epsilon_{\hbox{\sevenrm inj}}}

{\catcode`\@=11 
  \gdef\SchlangeUnter#1#2{\lower2pt\vbox{\baselineskip 0pt\lineskip0pt 
  \ialign{$\m@th#1\hfil##\hfil$\crcr#2\crcr\sim\crcr}}}} 
\def\gtrsim{\mathrel{\mathpalette\SchlangeUnter>}} 
\def\lesssim{\mathrel{\mathpalette\SchlangeUnter<}} 
\def\aa{{Astron. Astrophys.}}

\def\apj{ApJ}
\def\apjl{ApJ}
\def\apjsupp{ApJ Supp.}
\def\nat{Nature}
\def\app{Astroparticle Phys.}                   % DO NOT DELETE
\def\apss{Astr. Space Sci.}                     % DO NOT DELETE
\def\asr{Adv. Space Res.}                       % DO NOT DELETE
\def\mnras{{M.N.R.A.S.}}
\def\prl{Phys. Rev. Lett.}                      % DO NOT DELETE
\def\prd{Phys. Rev. D}                          % DO NOT DELETE
\def\ssr{Space Sci. Rev.}                       % DO NOT DELETE
\newcommand{\vol}[2]{$\;$\bf #1\rm , #2.}

\begin{document}

\title*{Lepton Acceleration in Pulsar Wind Nebulae}
%\titlerunning{Lepton Acceleration in Pulsar Wind Nebulae}

% 
\author{Matthew G. Baring}
\institute{Matthew G. Baring \at Department of Physics and Astronomy, MS-108,
                      Rice University, P. O. Box 1892, \\
                      Houston, TX 77251-1892, USA. e-mail: {\it baring@rice.edu} }

\maketitle

\begin{flushright}
\phantom{p}
\vspace{-250pt}
     To appear in Proc. of the inaugural ICREA Workshop on \\
     ``The High-Energy Emission from Pulsars and their Systems'' (2010),\\
     eds. N. Rea \& D. Torres, (Springer Astrophysics \& Space Science series).
\vspace{180pt}
\end{flushright}

\abstract{Pulsar Wind Nebulae (PWNe) act as calorimeters for the
relativistic pair winds emanating from within the pulsar light cylinder.
Their radiative dissipation in various wavebands is significantly
different from that of their pulsar central engines: the broadband
spectra of PWNe possess characteristics distinct from those of pulsars,
thereby demanding a site of lepton acceleration remote from the pulsar
magnetosphere. A principal candidate for this locale is the pulsar wind
termination shock, a putatively highly-oblique, ultra-relativistic MHD 
discontinuity.  This paper summarizes key characteristics of relativistic shock
acceleration germane to PWNe, using predominantly Monte Carlo simulation
techniques that compare well with semi-analytic solutions of the
diffusion-convection equation.  The array of potential spectral indices
for the pair distribution function is explored, defining how these
depend critically on the parameters of the turbulent plasma in the shock
environs.  Injection efficiencies into the acceleration process are also
addressed.  Informative constraints on the frequency of particle
scattering and the level of field turbulence are identified using the
multiwavelength observations of selected PWNe.  These suggest that the
termination shock can be comfortably invoked as a principal injector
of energetic leptons into PWNe without resorting to unrealistic
properties for the shock layer turbulence or MHD structure.}

\section{Introduction}
\label{sec:intro}
Pulsar wind nebulae (PWNe) have fascinated astronomers ever since the
discovery of the Crab Nebula. This source provides the template for PWN
studies because of the excellent multiwavelength spectral information
(de Jager \& Harding 1992; Atoyan \& Aharonian, 1996; Abdo et al. 2010b)
and stunning spatial imaging afforded by radio (historic), optical
(Hester et al. 1995 for {\it Hubble}) and X-ray (Weisskopf, et al. 2000
for {\it Chandra}) observations.  Its unparalleled observational quality
is driven by the exceptional powerhouse at its center, the high
spin-down power Crab pulsar.  The central pulsar fuels the dissipation
in its surrounding PWN (exemplified in the seminal Rees \& Gunn, 1974,
and Kennel \& Coroniti, 1984, models), with the nebula serving in a
symbiotic relationship as the calorimeter for the pulsar over its entire
spin-down history. Therefore, the interface between the central engine
and the nebula must play a principal role in setting up the emission seen
in PWNe.  This boundary is the {\bf pulsar wind termination shock}
(PWTS), where the wind is abruptly slowed by the ram pressure of the
circumstellar material; it forms the focus of this perspective on
lepton acceleration in PWNe.

This shock is a natural site for the acceleration of particles that
spawn the non-thermal radiation in PWNe that we observe.  It should
possess turbulent electrodynamic fields that can energize and
stochastically diffuse charges extremely efficiently.  Yet, the PWTS is
not the only possible site for leptonic acceleration.  Field
reconnection in and near the quasi-equatorial current sheet between the
pulsar light cylinder and the termination shock is an alternative (e.g.
Lyubarsky \& Kirk 2001; Kirk \& Skj\ae raasen 2003; P\'etri \& Lyubarsky
2007).  In compact regions such as X-points in the striped wind,
magnetic reconnection can release large amounts of energy as field
tension is converted to heat of particles.  Such a prospect needs fuller
exploration from a theoretical standpoint.  Reconnection theory needs to
make robust predictions of distributions and injection efficiencies
(from thermal gas) of accelerated populations in order to connect
effectively to PWNe observations.  The understanding of shock
acceleration is more developed in this regard, and accordingly is the
focus of this paper.   We note that the solar corona may prove a
powerful testing ground for honing models of reconnection in the same
way that the solar wind has demonstrated the general viability of
diffusive acceleration at non-relativistic shocks.  It should also be
remarked that all escaping pulsar wind leptons impact the surface of the
termination shock, whereas perhaps only a minority of such thread the
environs of the current sheet reconnection region.  Notwithstanding,
reconnection in the near wind zone may contribute significantly to the
evolution of the global MHD structure and associated wind parameters, as
well as generate some pre-acceleration, both of which in turn influence
the cumulative contribution of the PWTS as an injector to a pulsar wind
nebula over its active lifetime.

This paper summarizes the key aspects of diffusive acceleration at
relativistic shocks in general, and pulsar wind termination shocks in
particular.  As the injector of ultra-relativistic leptons, and ions,
into PWNe, this process is only indirectly probed by radiation
observations of nebulae.  The volumetric extension of PWNe encompasses
significant spatial stratification of both the nebular magnetic field
and the fluid flow speed, the model template for which is the
spherically symmetric Kennel \& Coroniti (1984) contribution.  Moreover,
temporal evolution is significant, with high energy electrons cooling
rapidly over the lifetime of a PWN like the Crab, driving synchrotron
``burn-off'' that is probed in the X-rays (see the review of Gaensler \&
Slane, 2006, for an extensive discussion of PWN observations and guiding
interpretative material). Yet multiwavelength coverage, from radio to
X-ray to high energy gamma-rays provides substantial constraints on the
PWTS acceleration process.  Presuming nebular fields in the range of
\teq{B\sim 0.1}mG implies pair Lorentz factors in the range
\teq{\gamma_e\sim 10^9 - 10^{10}} for the Crab to enable $\gamma$-ray
synchrotron emission.  Since the pulsar is unlikely to generate such
energetic particles, this demands efficient acceleration at the PWTS or
elsewhere. Other PWNe impose similar requirements.  The PWTS energy
budget divides into three components: (i) thermal downstream heat, (ii)
turbulent fields, and (iii) non-thermal shock-accelerated leptons, and
perhaps ions.  The balance between these is not yet fully understood,
though indications from plasma simulations are that these components are
not widely disparate in their energy densities.  The central
acceleration issue for PWN studies is whether a quasi-perpendicular
termination shock can generate a sufficient injection efficiency \teq{\epsinj} 
and the right spectral index in different energy ranges.  In this paper, it
becomes evident that the index issue can be satisfied in global terms
using the current understanding of diffusive acceleration at
relativistic shocks, while more work is needed to address the injection
issue in a satisfactory manner. 

\section{Lepton Acceleration at Relativistic Shocks}
 \label{sec:rel_accel}

To understand the nature of relativistic lepton injection into the PWN,
it is insightful to explore the general nature of particle acceleration
at relativistic shocks.  The key characteristic that distinguishes
relativistic shocks from their non-relativistic counterparts is their
inherent anisotropy of the phase space distribution function
\teq{f(\hbox{\bf p})} at any position.  This is due to rapid convection
of particles through and downstream away from the shock, since particle
speeds \teq{v} are never much greater than the downstream flow speed
\teq{u_2\sim c/3}: particle distributions never realize isotropy in
either fluid or shock rest frames.  This renders analytic approaches
more complicated (Peacock 1981) than in non-relativistic systems.  Early
analytic offerings on particle acceleration at relativistic shocks
focused on solutions of the diffusion-convection differential equation
in the test-particle approximation (e.g., Kirk \& Schneider 1987a;
Heavens \& Drury 1988; Kirk and Heavens 1989). These generally
specialized to the limit of extremely small angle scattering (SAS, or
{\it pitch angle diffusion}).  In particular, the eigenfunction solution
technique of Kirk \& Schneider (1987a) was later successfully extended
by Kirk et al. (2000) to the specific case of parallel,
ultrarelativistic shocks, i.e. those with upstream fluid flow Lorentz
factors \teq{\Gamma_1\gg 1} in the shock rest frame.  Kirk et al.
demonstrated that as \teq{\Gamma_1\to\infty}, the accelerated particle
distribution power-law index \teq{\sigma} (for \teq{dN/dp\propto p^2
f(\hbox{\bf p}) \propto p^{-\sigma}}) asymptotically approached a
constant, \teq{\sigma\to 2.23}, a value realized when
\teq{\Gamma_1\gtrsim 10}.  This result has been popularly invoked in
astrophysics models of various sources, but is of very restricted
applicability, as will become evident below. While diffusion-convection
differential equation approaches are usually restricted to SAS that
would be applicable to particle transport in quasi-linear field
turbulence regimes, recently they have been generalized by Blasi \&
Vietri (2005) and Morlini, Blasi \& Vietri (2007) to incorporate large
angle deflections in MHD turbulence of larger amplitudes \teq{\delta
B/B}.  The operating definition of such large angle scattering (LAS) is
that the particle experiences momentum deflections on typical angles
\teq{\thetascatt \gtrsim 1/\Gamma_1} in interactions with MHD turbulence
in the shock environs.  Clearly, for ultra-relativistic shocks, LAS can
be realized with quite modest deflections.

A central limitation of these analytic methods is that they are
restricted to power-law regimes, which are only realized when there is
no preferred momentum scale, i.e. far above the thermal injection
momentum.  Therefore they provide no probes of the injection efficiency
\teq{\epsinj} (defined to be the fraction of particles by number residing in the
non-thermal tail of the distribution), how \teq{\epsinj} connects key shock 
environmental parameters, and therefore how it correlates to the non-thermal
distribution index \teq{\sigma}.  Hence the niche for Monte Carlo
techniques for modeling diffusive transport in shocks.  Such
complementary simulation approaches have been employed for relativistic
shocks by a number of authors, including test-particle analyses by Kirk
\& Schneider (1987b), Ellison, Jones \& Reynolds (1990), and Baring
(1999) for parallel, steady-state shocks, and extensions to include
oblique magnetic fields by Ostrowski (1991), Ballard \& Heavens (1992),
Bednarz \& Ostrowski (1998), Ellison \& Double (2004), Niemiec \&
Ostrowski (2004), Stecker, Baring \& Summerlin (2007) and Baring \&
Summerlin (2009).  The Monte Carlo method successfully reproduced the
asymptotic \teq{\Gamma_1\to\infty} index value of \teq{\sigma \approx
2.23} in work by different groups (Bednarz \& Ostrowski 1998; Baring
1999; Achterberg, et al. 2001; Ellison \& Double 2002). There are two
main types of Monte Carlo simulation on the market: those that inject
prescribed field turbulence to effect diffusion of charges (e.g.
Ostrowski 1991; Bednarz \& Ostrowski 1998; Niemiec \& Ostrowski 2004),
and those that describe the diffusion by phenomenological scattering
parameters (e.g. Ellison, Jones \& Reynolds 1990; Ellison \& Double
2004; Baring \& Summerlin 2009). It is this latter variety that will
form the focus in this exposition, because of its ability to survey the
parameter space of acceleration characteristics in an incisive fashion.

Before outlining the essentials of the Monte Carlo technique used to
generate many of the results presented here, it should be noted that
there is a third popular approach to modeling particle acceleration at
relativistic shocks: full plasma or particle-in-cell (PIC) simulations
(e.g. Hoshino, et al. 1992; Nishikawa, et al. 2005; Medvedev, et al.
2005; Spitkovsky 2008).  PIC codes compute fields generated by mobile
charges, and the response of the charges to the dynamic electromagnetic
fields. Accordingly they are rich in their information on shock-layer
electrodynamics and turbulence, but pay the price of intensive demands
on CPUs.   This presently limits them to exploration of thermal and
suprathermal energies, so that full plasma simulations generally exhibit
largely Maxwellian distributions (Hoshino, et al. 1992; Nishikawa et al.
2005; Medvedev, et al. 2005). However, we note the isolated recent
suggestion (Spitkovsky 2008; Martins et al. 2009; Sironi \& Spitkovsky 2009) 
of non-thermal tails spanning relatively limited range of energies, 
generated by diffusive transport in PIC simulations, with the
thermal population still dominating the high-energy tail by number.  To
interface with astrophysical spectral data, a broad dynamic range in
momenta is desirable, and this is the natural niche of Monte Carlo
simulation techniques.

\subsection{The Monte Carlo Method}
 \label{sec:MC_technique}

As informative background to the ensuing results on relativistic planar
shocks, the structure of the simulation used to calculate diffusive
acceleration is now described.  It is a kinematic Monte Carlo technique
that has been employed extensively in supernova remnant and heliospheric
contexts, and is described in detail in numerous papers (e.g. Ellison,
Jones and Reynolds, 1990, hereafter EJR90; Jones \& Ellison 1991;
Ellison \& Double 2004; Baring \& Summerlin 2009).   It is conceptually
similar to Bell's (1978) test particle approach to diffusive shock
acceleration, and essentially solves a Boltzmann transport equation for
arbitrary orientations of the large scale MHD field {\bf B}. The
background fields and fluid flow velocities on either side of the shock
are uniform, and the transition at the shock is defined by the standard
relativistic MHD Rankine-Hugoniot conservation relations (e.g. Double et
al. 2004) that depend on both the sonic and Alfv\'enic Mach numbers. 
Particles are injected upstream of the shock with a Maxwell-Boltzmann
distribution of finite temperature, and convect and gyrate in the
laminar electromagnetic field, with their trajectories being governed by
a relativistic Lorentz force equation in the frame of the shock.  The
upstream fluid frame magnetic field is inclined at an angle
\teq{\thetaBone} to the shock normal. Because the shock is moving with a
velocity {\bf u}({\bf x}) relative to the plasma rest frame, there is,
in general, a {\bf u $\times$ B} electric field in addition to the bulk
magnetic field. Particle interactions with Alfv\'{e}n wave and other
hydromagnetic turbulence is modeled by using a phenomenological
scattering of the charges in the rest frame of the plasma.  The
scattering precipitates spatial diffusion of particles along magnetic
field lines, and to a varying extent, across them as well.  The
scatterings are also assumed to be quasi-elastic, an idealization that
is usually valid because in most astrophysical systems the flow speed
far exceeds the Alfv\'{e}n speed, and contributions from stochastic
second-order Fermi acceleration are small. The diffusion permits a
minority of particles to transit the shock plane numerous times, gaining
energy with each crossing via the coherent shock drift and diffusive
first-order Fermi processes.

A continuum of scattering angles, between large-angle or small-angle
cases, can be modeled by the simulation.  In the local fluid frame, the
time, \teq{\delta t_f}, between scatterings is coupled (EJR90) to the
mean free path, \teq{\lambda}, and the maximum scattering (i.e. momentum
deflection) angle, \teq{\thetascatt} via \teq{ \delta t_f\approx
\lambda\thetascatt^{2}/(6v)} for particles of speed \teq{v\approx c}. 
Here the mean fee path is that for turning the particles around along
field lines.  Usually \teq{\lambda} is assumed to be proportional to a
power of the particle momentum \teq{p} (see EJR90 and Giacalone, Burgess
and Schwartz, 1992, for microphysical justifications for this choice),
and for simplicity it is presumed to scale as the particle gyroradius,
\teq{r_g}, i.e. \teq{\lambda=\eta r_g\propto p}. Simulation results are
fairly insensitive to this choice.  Moreover, the scattering law is generally assumed
to be identical in both the upstream and downstream fluids.  Departures
from this can easily be accommodated, but usually incur only a change in the spatial 
scales for diffusion either side of the shock.   The parameter \teq{\eta} in the
model is a measure of the level of turbulence present in the system,
coupling directly to the amount of cross-field diffusion, such that
\teq{\eta =1} corresponds to the isotropic {\it Bohm diffusion} limit,
where the field fluctuations satisfy \teq{\delta B/B\sim 1}.  In the quasi-linear 
regime, \teq{\delta B/B\ll 1}, one expects that \teq{\eta} should scale
inversely as the variance of the field, i.e. \teq{\eta\propto (\delta B/B)^{-2}}.
In kinetic theory, \teq{\eta} couples the parallel (\teq{\kappa_{\parallel}=\lambda
v/3}) and perpendicular (\teq{\kappa_{\perp}}) spatial diffusion
coefficients via the relation
\teq{\kappa_{\perp}/\kappa_{\parallel}=1/(1+\eta^{2})} (Forman, Jokipii
\& Owens 1974; Ellison, Baring \& Jones 1995). In parallel shocks, where
the {\bf B} field is directed along the shock normal
(\teq{\thetaBone=0}), \teq{\eta} has only limited impact on the
resulting energy spectrum, principally determining the frequency of
scattering and hence the diffusive spatial scale normal to the shock. 
However, in oblique relativistic shocks where \teq{\thetaBone > 0}, the
diffusive transport of particles across the field (and hence through the
shock) becomes critical to retention of them in the acceleration
process. Accordingly, for such systems, the interplay between the field
angle and the value of \teq{\eta} controls the spectral index of the
particle distribution (Ellison \& Double 2004; Baring 2004), a feature
that is central to the interpretation of PWN spectra.

It should be remarked that this phenomenological description of
diffusion in Monte Carlo techniques is most appropriate at high energies
(where it is more or less commensurate with results from Monte Carlo
codes that inject prescribed turbulence), and omits the details of
microphysics present in plasma simulations such as PIC codes.  In the
injection domain at slightly suprathermal energies, the influences of
complex turbulent and coherent electrodynamic effects become important,
and will substantially modify the picture from that of pure diffusion
that is presented here; such is the niche of PIC simulations. Note also
that all subsequent simulation results presented here are obtained in
the {\it test particle approximation}, where the accelerated population
is not permitted to modify the overall MHD shock structure.

\subsection{Results for Relativistic Shock Acceleration}
 \label{sec:results}

\begin{figure}[t]
\includegraphics[scale=0.65]{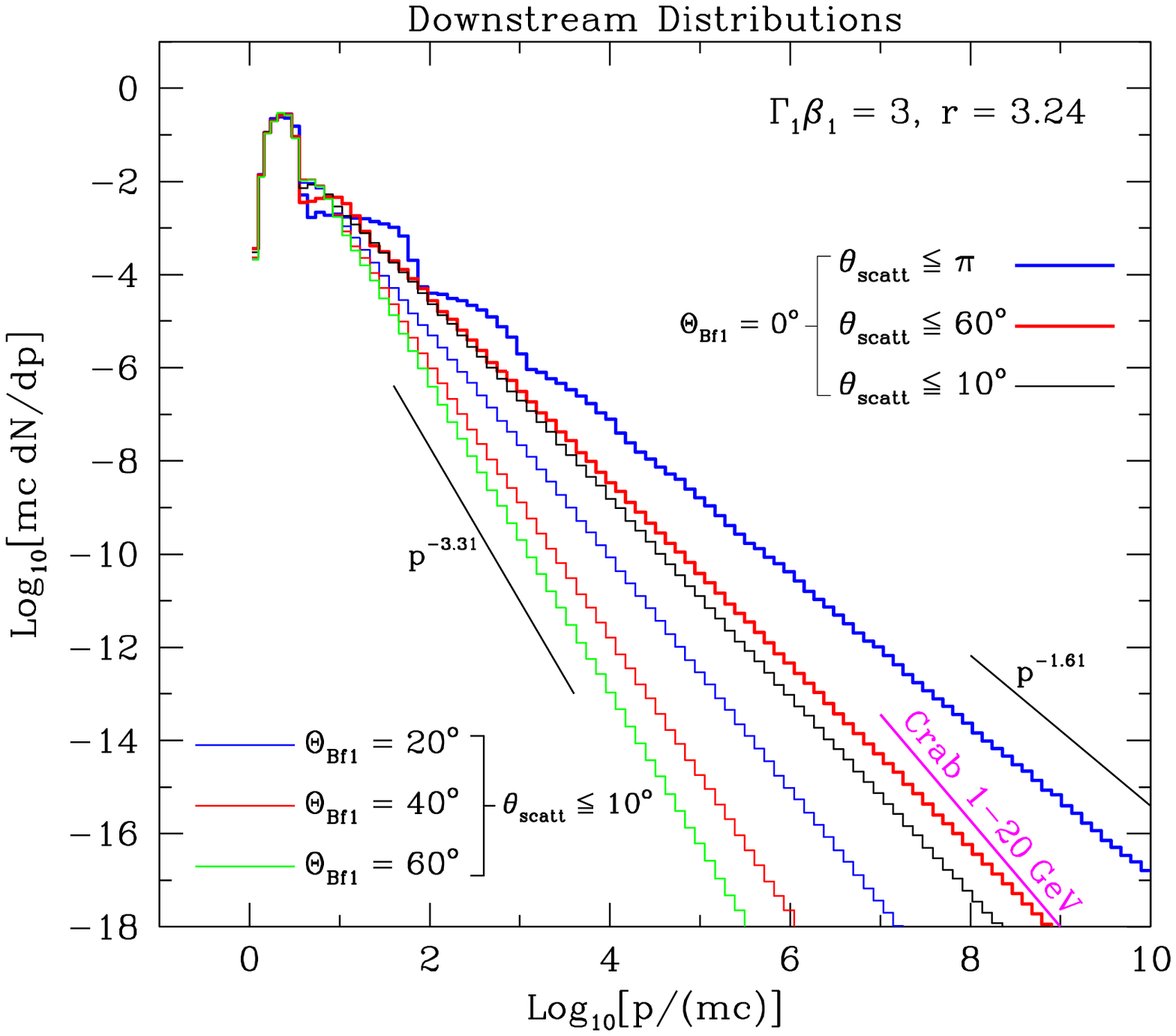} 
\caption{
Particle distribution functions  \teq{dN/dp} from mildly-relativistic
shocks (\teq{\Gamma_1\beta_1=3}, i.e. \teq{\beta_1=u_1/c=0.949}) of
upstream-to-downstream  velocity compression ratio
\teq{r=u_{1x}/u_{2x}\approx 3.24}. Simulation results can be divided
into two groups: parallel shock runs (\teq{\thetaBone=0^{\circ}}, upper
three histograms), and oblique, superluminal shock cases
(\teq{\thetaBone=20^{\circ}, 40^{\circ}, 60^{\circ}}, lower three
histograms).  Scattering off hydromagnetic turbulence was modeled by
randomly deflecting particle momenta by an angle within a cone, of
half-angle \teq{\thetascatt},  whose axis coincides  with the particle
momentum prior to scattering; the ratio of the diffusive mean free path
\teq{\lambda} to the gyroradius \teq{r_g} was fixed at \teq{\eta=\lambda
/r_g=5}. The heavyweight lines (two uppermost histograms) are for the
large angle scattering cases  (LAS: \teq{1/\Gamma_1\ll
\thetascatt\leq\pi}).  All other cases constitute pitch angle diffusion 
(small angle scattering: SAS) runs, when \teq{\thetascatt\ll 1/\Gamma_1} 
and the distributions become independent of the choice of \teq{\thetascatt}. 
All distributions asymptotically approach power-laws \teq{dN/dp\propto
p^{-\sigma}} at high energies.  For the two cases bracketing the results
depicted, the power-laws are indicated by lightweight lines, with
indices of \teq{\sigma=1.61} (\teq{\thetaBone=0^{\circ}},
\teq{\thetascatt\leq\pi}) and \teq{\sigma =3.31}
(\teq{\thetaBone=60^{\circ}}, \teq{\thetascatt\leq 10^{\circ}}),
respectively.  Also displayed is an indication of the index required to 
match {\it Fermi}-LAT \teq{> 1}GeV observations for the Crab Nebula, 
assuming uncooled inverse Compton emission. 
}
\label{fig:1}
\end{figure}

Representative particle differential distributions \teq{dN/dp\propto p^2
f(\hbox{\bf p})} that result from the simulation of diffusive
acceleration at mildly-relativistic shocks are depicted in
Figure~\ref{fig:1} (adapted from Baring 2009); the reader can survey
Ellison \& Double (2004), and Stecker, Baring and Summerlin (2007,
hereafter SBS07) for \teq{\Gamma_1\gg 1} simulation results that possess
similar character to the parallel shock (\teq{\thetaBone =0^{\circ}})
examples in the Figure. These distributions are obtained just downstream 
of the shock and are measured in the shock rest frame.  They are equally 
applicable to electrons or ions, and so the mass scale is not specified; 
presuming that the wind loss from pulsars is dominated by pairs, the mass 
scale is nominally \teq{m_e}. A striking feature is that the slope and shape 
of the non-thermal particle distribution depends on the nature of the
scattering.  The often cited asymptotic, ultrarelativistic index of
\teq{\sigma =2.23} for \teq{dN/dp\propto p^{-\sigma}} mentioned above is
realized only for parallel shocks with \teq{\thetaBone =0^{\circ}} in
the mathematical limit of small (pitch) angle diffusion (SAS), where the
particle momentum is stochastically deflected on arbitrarily small
angular (and therefore temporal) scales. As mentioned above, in
practice, SAS results when the maximum scattering angle
\teq{\thetascatt} is inferior to the Lorentz cone angle \teq{1/\Gamma_1}
in the upstream region. In such cases, particles diffuse in the region
upstream of the shock only until their velocity's angle to the shock
normal exceeds around \teq{1/\Gamma_1}, after which they are rapidly
swept downstream of the shock. The Figure indicates clearly that when
the field obliquity \teq{\thetaBone} increases, so also does the index
\teq{\sigma}, with values greater than \teq{\sigma\sim 3} arising for
\teq{\thetaBone\gtrsim 50^{\circ}} for this mildly-relativistic
scenario.  This is a consequence of more prolific convection downstream
away from the shock.

Figure~\ref{fig:1} also shows results for large angle scattering
scenarios (LAS, with \teq{4/\Gamma_1\lesssim \thetascatt\lesssim\pi}),
where the distribution is highly structured and much flatter on average
than \teq{p^{-2}}.  The structure becomes more pronounced for large
\teq{\Gamma_1} (see Baring 2004; Ellison \& Double 2004; SBS07, for
details), and is kinematic in origin, where large angle deflections lead
to fractional energy gains between unity and \teq{\Gamma_1^2} in
successive shock crossings.  Each structured bump or spectral segment
corresponds to an increment of two in the number of shock transits
(Baring 2004).  For \teq{p\gg mc}, they asymptotically relax to a
power-law, in this case with index \teq{\sigma\approx 1.61}.
Intermediate cases are also depicted in Figure~\ref{fig:1},  with
\teq{\thetascatt\sim  4/\Gamma_1}.  The spectrum is  smooth, like for
the SAS case, but the index is lower than 2.23.  From the plasma physics
perspective, magnetic turbulence could easily be sufficient  to effect
scatterings on this intermediate angular scale, a contention that
becomes even more germane for ultrarelativistic shocks (SBS07). Note
that there is a clear trend (e.g. see EJR90; Baring 2004; SBS07) of
declining \teq{\sigma} for higher \teq{\Gamma_1}, the consequence of an
increased kinematic energy boosting in collisions with turbulence.

The plot in Figure~\ref{fig:1} includes an indication of the particle
distribution index required to match the {\it Fermi} observations of the
Crab Nebula. The {\it Fermi}-LAT spectral index in the 1--20 GeV range,
corresponding to a putative inverse Compton signal, is
\teq{\alpha_\gamma =1.64} (see Abdo et al. 2010a, and specifically
Figure 5 therein). In the case where this corresponds to the {\it in
situ} accelerated population (i.e. the population is uncooled on the
relevant timescales), one finds that \teq{\sigma = 2\alpha_{\gamma}-1 =
2.28}.  Such a scenario is depicted by the ``Crab 1--20 GeV" line in
Fig.~\ref{fig:1}. In contrast, if inverse Compton cooling is
sufficiently rapid as to define the total {\it Fermi} \teq{>1}GeV
spectrum, then \teq{\sigma = 2\alpha_{\gamma}-2 = 1.28}. Thus,
strongly-cooled IC models would suggest large angle scattering is active
in the Crab Nebula termination shock, if it is {\it superluminal}.

\begin{figure}[t]
\includegraphics[scale=0.65]{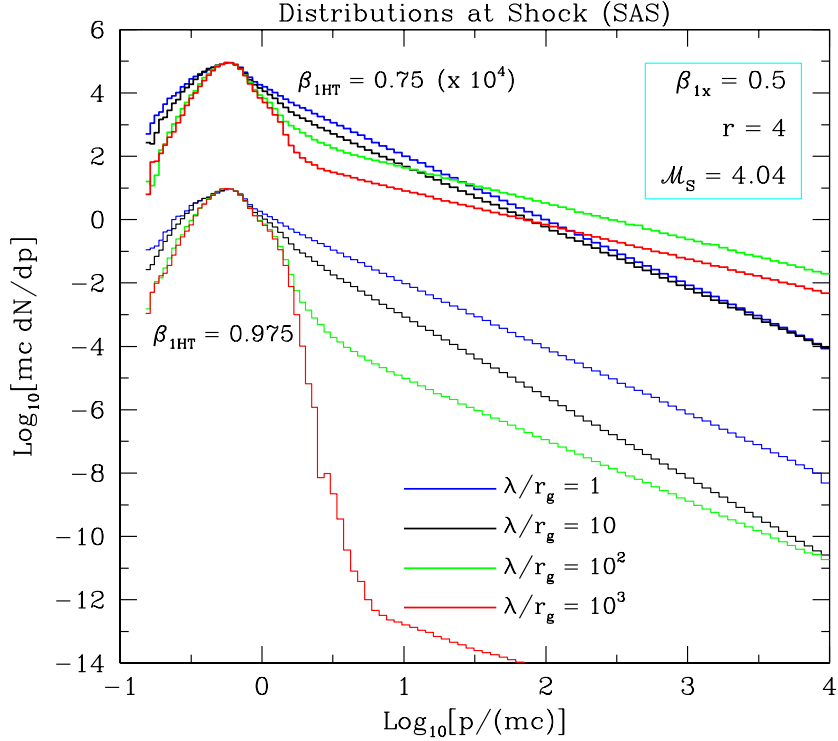}
\caption{Particle distribution functions  \teq{dN/dp} from mildly-relativistic
sub-luminal shocks (\teq{\Gamma_{1x}\beta_{1x}=0.577}, i.e.
\teq{\beta_{1x}=u_{1x}/c=0.5}) of upstream-to-downstream  velocity
compression ratio \teq{r=u_{1x}/u_{2x}\approx 4}.  Simulation results
are depicted for two upstream fluid frame magnetic field obliquities,
labelled by their corresponding de Hoffman-Teller frame upstream flow
speeds \teq{\beta_{\hbox{\sevenrm 1HT}} = \beta_{1x}/\cos\thetaBone}.
These are in distinct groups of four: \teq{\thetaBone=48.2^{\circ}}
(\teq{\beta_{\hbox{\sevenrm 1HT}} = 0.75}, multiplied by \teq{10^4}) for the
upper four histograms, and \teq{\thetaBone=59.1^{\circ}}
(\teq{\beta_{\hbox{\sevenrm 1HT}} = 0.975}) for the lower four
histograms. Scattering off hydromagnetic turbulence was modeled by
randomly deflecting particle momenta by an angle within a cone, of
half-angle \teq{\thetascatt},  whose axis coincides  with the particle
momentum prior to scattering; four different ratios of the diffusive
mean free path \teq{\lambda} to the gyroradius \teq{r_g} were adopted
for each \teq{\thetaBone}. All results were for small angle scattering
(SAS), when \teq{\thetascatt\ll 1/\Gamma_1} and the distributions
become independent of the choice of \teq{\thetascatt}. A low sonic Mach
number \teq{\machson} was chosen so as to effectively maximize the efficiency of
injection from thermal energies.  Adapted from Baring \& Summerlin (2009).
}
\label{fig:2}
\end{figure}

Now for an important definition pertaining to the following discussion.
The MHD phase space of relativistic shocks bifurcates neatly into two
regimes.  In general, Monte Carlo simulations ``operate'' in a shock
rest frame named the normal incidence frame (NIF), where the upstream
flow is directed along the shock normal (usually chosen to be the
\teq{x}-direction, a convention adopted here).  In this frame, the
upstream magnetic field is inclined to shock normal by an angle of
\teq{\thetaBnone}.  Due to relativistic aberration effects, generally
\teq{\thetaBnone\neq\thetaBone}, with equality arising only in truly
non-relativistic shocks. For many systems, there is also a shock rest
frame called the de Hoffman-Teller (HT) frame (identified by de Hoffman
\& Teller 1950), which is obtained by a boost \teq{\uHTone\equiv
\betaHTone c= u_{1x}/\cos\thetaBone} along the magnetic field so as to
bring the shock to rest.  In this HT frame, there are no static electric
fields, implying no {\bf E} \teq{\times} {\bf B} drifts parallel to the
shock plane. {\it Subluminal} shocks are defined to be those where the
HT flow speed \teq{\betaHTone} corresponds to a physical speed, less
than unity, i.e. the upstream field obliquity satisfies
\teq{\cos\thetaBone < \beta_{1x} \equiv u_{1x}/c}.  When \teq{\betaHTone
> 1}, the de Hoffman-Teller frame does not exist, and the shock is said
to be {\it superluminal}.  This division naturally demarcates a
dichotomy for the gyrational characteristics of charges orbiting in the
shock layer.  Subluminal shocks permit many gyrational encounters of
charges with the shock interface, and therefore also reflection of them 
into the upstream region. This implies efficient trapping (e.g see
Baring \& Summerlin 2009), and effective acceleration. In contrast, for
superluminal shocks, in the absence of deflections of particles by
magnetic turbulence,  the convective power of the flow compels particles
to rapidly escape downstream (e.g. Begelman \& Kirk 1990), thereby
suppressing acceleration.  In such cases, particles sliding along the
magnetic field lines would have to move faster than the speed of light
in order to return to the upstream side of the shock.  Such dramatic
losses from the acceleration mechanism can only be circumvented by
strong cross field diffusion precipitated by large amplitude field
turbulence fields (e.g. Jokipii 1987; Ellison, Baring \& Jones 1995),
i.e. essentially close to the Bohm limit.

The focus now turns to displaying the array of expectations for
subluminal relativistic shocks.  Principally, we will explore how the
distribution index \teq{\sigma} and injection efficiency depend on the
effective frequency \teq{\lambda/r_g} of scatterings, and the upstream
field obliquity \teq{\thetaBone}.  Representative particle (lepton or
ion) differential distributions \teq{dN/dp} that result from the
simulation of diffusive acceleration at mildly-relativistic shocks of
speed \teq{\beta_{1x}=0.5} are depicted in Figure~\ref{fig:2}.  These
distributions were generated for \teq{\thetascatt \lesssim 10^\circ},
i.e. in the SAS regime. Results are displayed for two different upstream
fluid frame field obliquities, namely \teq{\thetaBone=48.2^{\circ}} and
\teq{\thetaBone=59.1^{\circ}}, with corresponding de Hoffman-Teller
frame dimensionless speeds of \teq{\betaHTone =\beta_{1x}/\cos\thetaBone
= 0.75} and \teq{0.975}, respectively. The distributions clearly exhibit
an array of indices \teq{\sigma}, including very flat power-laws, that
are not monotonic functions of either the field obliquity
\teq{\thetaBone} or the key diffusion parameter \teq{\eta =\lambda
/r_g}. Fig.~\ref{fig:2} also emphasizes that the normalization of the
power-laws relative to the low momentum thermal populations (and hence 
the injection efficiency \teq{\epsinj}) is a
strongly-declining function of \teq{\lambda /r_g}.  Quantitatively, \teq{\epsinj}
drops from \teq{0.1 - 0.2} in the Bohm limit cases to less than \teq{10^{-4}} 
for \teq{\lambda/r_g=10^2} when \teq{\betaHTone = 0.975}.  This is a direct
consequence of a more prolific convection of suprathermal particles
downstream of the shock that suppresses diffusive injection from thermal
energies into the acceleration process.  Such losses are even more
pronounced when \teq{\lambda /r_g \geq 10^4}, to the point that
acceleration is not statistically discernible for \teq{\betaHTone >
0.98} runs with \teq{10^4} simulated particles. This property is salient
for the pulsar wind nebula context discussed below.

\begin{figure}[t]
\includegraphics[scale=0.65]{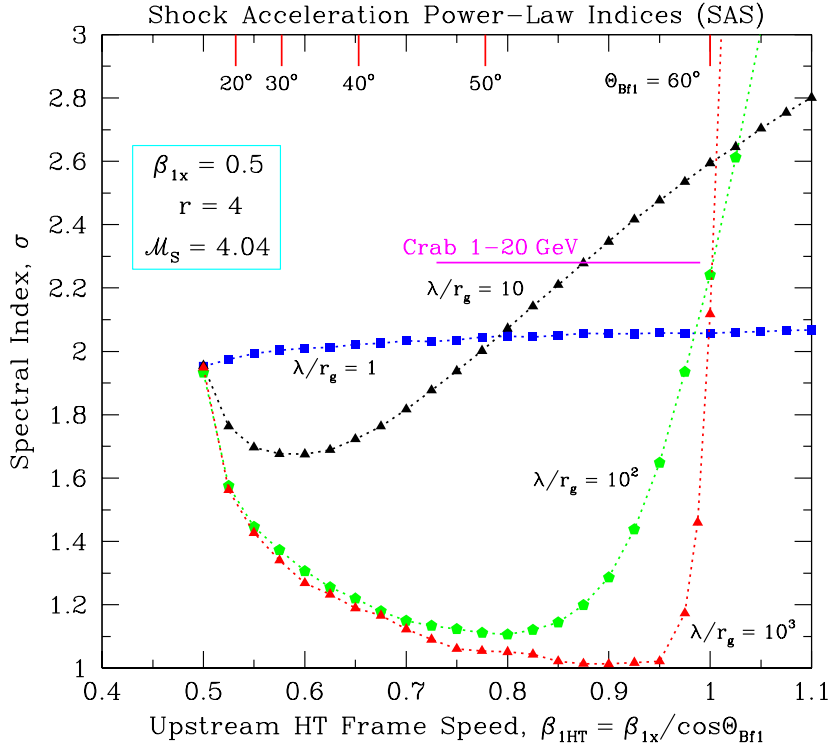}
\caption{Power-law indices \teq{\sigma} for simulation runs in the limit of 
small angle scattering (pitch angle diffusion), for mildly-relativistic shocks
of upstream flow speed \teq{\beta_{1x}\equiv u_{1x}/c =0.5}, and an MHD
velocity compression ratio \teq{r=4}. The indices are displayed as
functions of the effective de Hoffman-Teller frame upstream flow speed
\teq{\beta_{\hbox{\sevenrm 1HT}} = \beta_{1x}/\cos\thetaBone}, with
select values of the fluid frame field obliquity \teq{\thetaBone} marked
at the top of the panel. The displayed simulation index results were
obtained for different diffusive mean free paths \teq{\lambda} parallel
to the mean field direction, namely \teq{\lambda/r_g=1} (squares),
\teq{\lambda/r_g=10} (triangles), \teq{\lambda/r_g=10^2} (pentagons),
and \teq{\lambda/r_g=10^3} (triangles), as labelled.  The short
heavyweight line indicates the approximate spectral index
\teq{\sigma} that is appropriate to match {\it Fermi}-LAT \teq{> 1}GeV 
observations for the Crab Nebula, assuming uncooled inverse Compton 
emission is operable.  Note that the indices for the \teq{\betaHTone =0.75}, 
\teq{0.975} cases correspond to those of the distributions exhibited 
in Fig.~\ref{fig:2}.
}
\label{fig:3}
\end{figure}

A parameter survey for diffusive acceleration at a typical
mildly-relativistic shock is exhibited in Figure~\ref{fig:3}, where only
the pitch angle diffusion limit was employed. The power-law index
\teq{\sigma} is plotted as a function of the de Hoffman-Teller frame
dimensionless speed \teq{\betaHTone =\beta_{1x}/\cos\thetaBone}.  It is
clear that there is a considerable range of indices \teq{\sigma}
possible for non-thermal particles accelerated in mildly relativistic
shocks. A feature of this plot is that the dependence of \teq{\sigma} on
field obliquity is non-monotonic.  When \teq{\lambda /r_g\gg 1}, the
value of \teq{\sigma} at first declines as \teq{\thetaBone} increases
above zero, leading to very flat spectra.  As \teq{\betaHTone}
approaches and eventually exceeds unity, this trend reverses, and
\teq{\sigma} then rapidly increases with increasing shock obliquity. 
This is the character of near-luminal and superluminal shocks evident in
Fig.~\ref{fig:2}: it is caused by inexorable convection of particles
away downstream of the shock, steepening the distribution dramatically.
The only way to ameliorate this rapid decline in the acceleration
efficiency is to reduce \teq{\lambda /r_g} to values below around
\teq{10}.  Physically, this corresponds to increasing the hydromagnetic
turbulence to high levels that force the particle diffusion to approach
isotropy.  This renders the field direction immaterial, and the shock
behaves much like a parallel, subluminal shock in terms of its diffusive
character.  Charges can then be retained near the shock for sufficient
times to accelerate and generate suitably flat distribution functions. 
This defines a second core property illustrated in Fig.~\ref{fig:3}:
\teq{\sigma} is only weakly dependent on \teq{\thetaBone} when
\teq{\lambda /r_g < 10}. Observe that the indication of the particle
distribution index corresponding to {\it Fermi}-LAT observations in the
1--20 GeV range (for uncooled inverse Compton models; same as in
Fig.~\ref{fig:1}) suggests low values of \teq{\lambda /r_g} and
proximity of the shock obliquity to the subluminal/superluminal
boundary.  This inference will be developed further below.

It is appropriate to identify briefly the reason why the distribution
indices approach \teq{\sigma\sim 1} for subluminal shocks when
\teq{\lambda/r_g\gg 1}, i.e. the field is almost laminar. The origin of
the extremely flat distributions with \teq{\sigma\sim 1} is in the
coherent effect of {\it shock drift acceleration} at the shock
discontinuity, discussed extensively in Baring \& Summerlin (2009). This
phenomenon is due to the energy gain of charges when they repeatedly
encounter {\bf u}\teq{\times}{\bf B} electric fields (in frames other
than the HT frame) in gyrations straddling the shock discontinuity. 
Such gains are experienced between episodic upstream excursions as
charges more or less retain gyrophases that permit reflection from the
shock for long periods of time.  Reducing \teq{\lambda /r_g}, and
thereby introducing extremely modest amounts of cross-field diffusion,
disrupts this coherence, removes particles from the shock layer, and
steepens the spectrum. It is not clear that astrophysical relativistic
shocks can contain such low levels of turbulence as to access this
academically interesting regime of phase space.

In concluding this overview of particle acceleration characteristics at
relativistic shocks, it is noted that the results from these Monte Carlo
simulations are in good agreement with those from other techniques, such
as semi-analytic numerical solutions of the diffusion-convection
equation, and also other Monte Carlo research initiatives.  In
particular, the artificially high choice of the compression ratio
\teq{r=4} in Figures~\ref{fig:2} and~\ref{fig:3} was adopted to
facilitate comparison with the semi-analytic work of Kirk \& Heavens
(1989).  The reader is referred to Baring \& Summerlin (2009) and Baring
(2010) for more details on such simulation validation.

\section{The Quasi-Perpendicular Pulsar Wind Termination Shock}
 \label{sec:termshock}

The discussion now turns to lepton acceleration in pulsar wind
termination shocks. While it is clear that their upstream flow speeds
should be ultrarelativistic, it is unclear how fast they are.  The
historical paradigm of upstream bulk Lorentz factors \teq{\Gamma_1\sim
10^5} in the Crab Nebula has been promulgated from the seminal work of
Kennel \& Coroniti (1984). Pulsars can easily generate such bulk flows
propagating out through the light cylinder, since the accelerating
potentials in their gaps must energize primary electrons to at least
\teq{\gamma_e\sim 10^6-10^7}. This is true for both outer gap models
(e.g. Cheng, Ho \& Ruderman 1986; Romani 1996) or polar cap scenarios
(e.g. Daugherty \& Harding 1982; 1996) for the electromagnetic
dissipation zone in gamma-ray puslars.  If radiation reaction-limited
curvature emission is what is principally responsible for the GeV
emission seen in a host of {\it Fermi}-LAT pulsars (see the {\it Fermi}
pulsar catalog compendium in Abdo et al. 2010b), then one can simply
derive the relation \teq{\gamma_e^3\lambar /\rho_c\sim 2\emax /3} for
emission turnovers \teq{\emax\sim 5\times 10^3} (in units of
\teq{m_ec^2}) in the GeV band.  Here \teq{\rho_c} is the magnetic field
curvature radius, which is some fraction of the light cylinder radius
\teq{\Rlc = Pc/(2\pi)} for pulsar period \teq{P} seconds.  Also,
\teq{\lambar =\hbar/(m_ec)=3.862\times 10^{-11}}cm is the electron
Compton wavelength over \teq{2\pi}.  With \teq{10^6\hbox{cm} < \rho_c <
10^9}cm, it is inferred that primaries assume Lorentz factors
\teq{10^6\lesssim \gamma_0 \lesssim 10^7} in a broad array of 
young to middle-aged pulsars.  

However, pair cascading is
rife in both the polar cap and slot gap/outer gap gamma-ray pulsar
pictures.  Much of the pair creation (magnetic one photon or
conventional two-photon) occurs outside the gaps containing accelerating
potentials. Several generations of pair production ensue, precipitating
large pair multiplicities \teq{\eta_{\pm}\sim 10 - 10^4} (e.g. see
Daugherty \& Harding 1982 for polar cap realizations, and Muslimov \&
Harding 2003 for slot gap results).  Furthermore, similar values are
obtained by De Jager (2007), who used the TeV inverse Compton flux in
PSR B1509-58 and PSR B1823-13 to infer the total electron deposition
integrated over the ages of their nebulae, thereby acting as a
calorimeter for their pulsar pair multiplicities (see also Bucciantini,
Arons and Amato 2010 for generally higher estimates for
\teq{\eta_{\pm}}). Simple energy conservation in the cascading process
trades multiplicity for Lorentz factor, so that most of the emergent
pairs propagating outwards from the gap region assume typical Lorentz
factors of \teq{\gamma_{\pm}\sim \gamma_0/\eta_{\pm}}. This then defines
fiducial bulk Lorentz factors for the pair flow escaping towards the distant
termination shock, so that for nebular modeling purposes
\teq{\Gamma_1\sim 10^2-10^4} may be more representative of the flow just
upstream of the PWTS than the higher Kennel \& Coroniti (1984) value.
However, we note that since the wind is strongly magnetically-dominated
at the light cylinder, mysteriously transitioning to a plasma-dominated flow
at the termination shock (the so-called infamous \teq{\sigma} problem),
conversion of Poynting flux to bulk plasma kinetic energy is a distinct 
possibility for raising the value of \teq{\Gamma_1}, perhaps taking
advantage of magnetic reconnection in and near the current sheet.

The obliquity of the PWTS is less subject to such debate.  If the
termination shock is a fairly regular spatial structure, it must be
highly oblique or an essentially perpendicular shock
(\teq{\thetaBone\sim 90^{\circ}}) in the equatorial wind zone, and also
at much higher pulsar latitudes.  Within the light cylinder, this
follows from the winding up of the field in a classic Parker spiral,
just like the solar wind termination shock (e.g. see Bogovalov, 1999,
for a discussion of MHD structure in oblique rotators).  Only directions
outside the pulsar polar regions can possess more radial fields that
permit the shock to be merely oblique, or even quasi-parallel. The
actual solid angle (centered on the pulsar) portion of the PWTS that is
quasi-perpendicular depends on the obliquity of the rotator, how the
virtually rigid inner magnetospheric field morphology causally maps
over to the field outside the light cylinder, and how the field geometry
is modified by plasma loading. Yet it is in all probability large,
regardless of whether \teq{\Gamma_1} is as high as \teq{10^5} or as low
as \teq{10^2}.  From the MHD simulations of the Crab pulsar wind of
Komissarov \& Lyubarsky (2004), it is clear the the termination shock is
non-spherical, being radially compressed in the polar zones. One can
also entertain the possibility that the termination shock is slightly
rippled, akin to what is an emerging paradigm for the solar wind shock
based on the surprising magnetometer and energetic particle data
acquired by the Voyager I and II spacecraft in the last few years.  This
can then permit localized regions of the PWTS to be subluminal or
marginally superluminal.  Or it can provide seeds for acceleration in a
perpendicular shock zone from remote, but merely oblique shock environs.
  However, observational support for any such a conjecture is a long way
off since it requires angular resolutions exceeding that of {\it Hubble}
and {\it Chandra} to probe such PWTS geometry in bright PWNe like the
Crab (see Hester et al. 1995 for {\it Hubble} and ROSAT images) and MSH
15-52 (see Gaensler et al. 2002 for {\it Chandra} imaging).

The content of the PWTS is generally presumed to be an electron-positron
pair plasma.  This derives from the leading models for dissipation in
the pulsar magnetosphere: pairs are rife therein due to the relative
ease of leptons being stripped from the neutron star surface. Thermionic
emission is possible in pulsars with higher surface temperatures.
Moreover, if sufficiently intense parallel electric fields persist in
the atmosphere, space-charge limited ion acceleration can proceed
(Ruderman \& Sutherland 1975; Arons and Scharlemann 1979).  Such a
prospect drove ideas that young neutron stars (Blasi, Epstein \& Olinto
2000) and magnetars (Arons 2003) could act as accelerating sources of
ultra-high energy cosmic rays. In the context of PWNe, baryonic loading
of the wind that impacts the termination shock is possible, and
inherently alters the character of the shock. Low energy charges then
become subject to cross-shock potentials in the shock layer, since the
inertial (i.e. gyrational) scales of the different species are widely
disparate (e.g. see the discussion in Baring \& Summerlin 2007). This
can act to redistribute the thermal energy of the charges, possibly
enhancing the injection and acceleration efficiency of leptons by
tapping the inertia of the incoming thermal ions. Even if the pulsar
wind is pair-dominated, it is still possible that the PWTS interface
picks up ions from the proximate hydrogenic ejecta and feeds them into
the acceleration process.  Observational constraints on hadronic
contributions to PWN gamma-ray emission are substantial.  For example,
the multi-zone models of multiwavelength emission in the Crab nebula of
Atoyan \& Aharonian (1996) indicate that pion decay emission from
PWTS-accelerated protons colliding with cold ambient hydrogen lies
comfortably below the inverse Compton signal in the 100 MeV -- 1 TeV
band, and is only likely to be detectable at energies \teq{>10}TeV.  The
flat spectrum and absence of any pion decay feature in the {\it
Fermi}-LAT spectrum of the Crab (Abdo et al. 2010a) strongly suggest
that the pair component of the PWTS is the most relevant.  Given that
both environmental and neutron star-driven baryonic loading are
uncertain, and the observational mandate for treating hadronic emission
in PWNe is limited, the discussion below will focus on pure lepton
models for wind nebulae.

\section{Connecting to PWN Observations}
 \label{sec:observe}

The emphasis now turns to making direct inferences on the pulsar wind
termination shock environment and its lepton acceleration
characteristics using the multiwavelength observations of nebular
emission.  This necessarily connects to the non-thermal power-law
distribution indices \teq{\sigma}.  For the best known and most
intensively-studied case of the Crab, the radio spectral index is quite
flat at \teq{\alpha_{\gamma}=1.26} (e.g. Wright et al. 1979), the X-ray
index is steeper at \teq{\alpha_{\gamma} \sim 2.1} (see Weisskopf, et
al. 2000; Atoyan \& Aharonian 1996), the 1--20 GeV $\gamma$-ray spectrum has
\teq{\alpha_{\gamma}=1.64} (e.g. Abdo et al. 2010a) which slowly breaks
to \teq{\alpha_{\gamma}\sim 2.5} above 1 TeV. The radio spectrum is not
flat enough for synchrotron self-absorption, and no low frequency
turnover that would be a signature of a minimum lepton Lorentz factor is
observed.  Accordingly, injection of pairs into the acceleration process must
take place at energies below around 3--10 GeV.  These characteristics
are more or less representative of other PWNe: the radio index generally
lies around 1.3 (see Gaensler \& Slane 2006), and is flatter than the
X-ray and TeV gamma-ray spectra, a nice synopsis of which is provided in
the recent review of Kargaltsev \& Pavlov (2010). Even if effective
radiative cooling is invoked at the maximum pair energies generating
X-ray synchrotron emission, it becomes evident from these properties
that the pair injection spectrum is {\it convex}, ranging from
\teq{\sigma\sim 1.5} below around 30 GeV to \teq{\sigma\sim 2.3} well
above 1 TeV (e.g. see Bucciantini, Arons \& Amato 2010). As will become
evident shortly, this is a significant constraint on diffusive
acceleration at the PWTS.

\begin{figure}[t]
\includegraphics[scale=0.65]{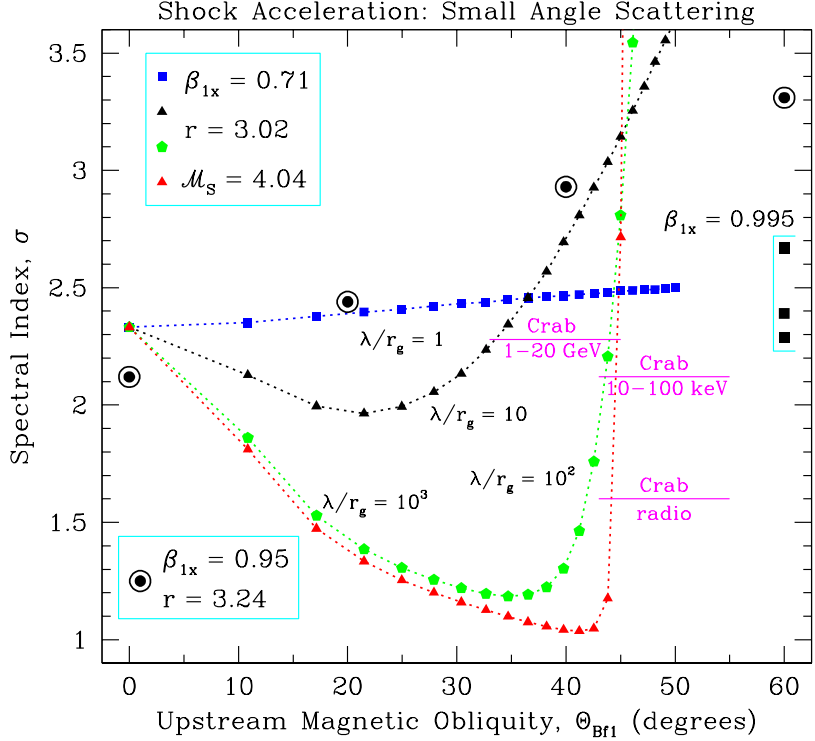}
\caption{
Power-law indices \teq{\sigma} for simulation runs in the limit of small
angle scattering, for relativistic shocks of three different speeds. The
indices are displayed as functions of the fluid frame field obliquity
\teq{\thetaBone} (contrasting Fig.~\ref{fig:3}).  Simulation data for
the points connected by dotted lines were for an of upstream flow speed
\teq{\beta_{1x}\equiv u_{1x}/c =0.71}, and an MHD velocity compression
ratio \teq{r=3.02}; for these runs, obliquities \teq{\thetaBone >
45^\circ} constitute superluminal shocks. These index results were
obtained for different diffusive mean free paths \teq{\lambda} parallel
to the mean field direction, namely \teq{\lambda/r_g=1} (squares),
\teq{\lambda/r_g=10} (triangles), \teq{\lambda/r_g=10^2} (pentagons),
and \teq{\lambda/r_g=10^3} (triangles), as labelled. Data for the higher
shock speed (\teq{\Gamma_1\beta_{1x}=3\Rightarrow \beta_{1x}\equiv
u_{1x}/c =0.949}) spectra displayed in Fig~\ref{fig:1} (SAS only) are
exhibited as circular points with dots centered therein. These are
mostly superluminal and corresponded to \teq{\lambda/r_g=5}. The final
subset of datapoints are the three filled squares grouped at
\teq{\thetaBone =60^{\circ}} for runs with \teq{\Gamma_1\beta_{1x}=10}
(\teq{\beta_{1x}\approx 0.995}) with \teq{r=3.02} (Summerlin \& Baring,
in preparation).  These were obtained for \teq{\lambda /r_g=1, 3, 6}
ranging from the bottom to the top. As with previous Figures, short
heavyweight lines are used to indicate the approximate spectral index
\teq{\sigma} that is appropriate to match Crab Nebula spectra in different 
wavebands: radio (presumed to be uncooled synchrotron), 10-100
keV X-ray (cooled synchrotron emission) and  {\it Fermi}-LAT \teq{>1}GeV 
observations (uncooled inverse Compton); see the text for a discussion.
}
\label{fig:4}
\end{figure}

We now identify the shock conditions required to generate these detected
spectral indices, by considering shocks of higher speeds than in the
previous Section.  Representative spectral index results from Monte
Carlo simulation runs are exhibited in Figure~\ref{fig:4} (from
Summerlin \& Baring, in preparation), the \teq{\beta_{1x}=0.71} portion
of which mirrors those presented in Figure~\ref{fig:3}. When the Bohm
limit of \teq{\lambda /r_g=1} is realized, the non-thermal distribution
index is approximately independent of the field obliquity.  When the
shock is superluminal, the index \teq{\sigma} is a rapidly increasing
function of \teq{\thetaBone}. In subluminal regimes due to the powerful
convective infleunces, when the field is laminar and \teq{\lambda
/r_g\gg 1}, very flat spectra can be realized because particles can be
trapped in the shock layer and shock drift acceleration is very
effective.  Note also, that inefficient injection from thermal energies
is then operative, as is exhibited in Fig.~\ref{fig:2}. The
\teq{\beta_{1x}=0.95} indices are those taken from Fig.~\ref{fig:1} and
indicate a moderate increase with obliquity in the superluminal regime. 
Such an increase is tempered relative to the \teq{\beta_{1x}=0.71},
\teq{\lambda /r_g=10} situation largely because \teq{\lambda /r_g} and
the compression ratio are higher.  The superluminal and
ultra-relativistic \teq{\Gamma_1\approx 10}, \teq{\thetaBone
=60^{\circ}} results are possibly the most representative of the PWTS.
They illustrate a significant sensitivity of \teq{\sigma} to
\teq{\lambda /r_g}, yet the indices are lower than those for the shocks
of lower speeds \teq{\beta_{1x}}.  This is caused by the increased
kinematic energy gains in shock crossings for high \teq{\Gamma_1} for
quasi-elastic interactions between charges and MHD turbulence in the
shock layer.

The spectral indices observed for the Crab nebula in radio, X-ray and
gamma-ray wavebands, as marked in Figure~\ref{fig:4}, offer clear
constraints on the shock environment, if diffusive acceleration at the
PWTS is the operable injection in PWNe. These can be taken to be more or
less representative of the broader population of PWNe, though variations
exist in the observational database.  The flat radio (synchotron)
spectra demand that the turbulence generate large mean free paths along
{\bf B} if the PWTS is subluminal.  Shocks by their nature generate
turbulence at levels that make this scenario unlikely (see the
discussion in Baring \& Summerlin 2009), disrupting the coherence that
permits shock drift acceleration to operate prolifically.  Given that
the PWTS is very probably superluminal over most of its surface, the
small angle scattering regime cannot supply flat enough acceleration
distributions.  Large angle scattering can though, as is evident in
Fig.~\ref{fig:1} and in Stecker, Baring \& Summerlin (2007). This is not
an unduly restrictive demand in ultra-relativistic shocks, since LAS is
delineated by deflections \teq{\thetascatt\gtrsim 1/\Gamma_1}, and it is
easy to envisage that MHD turbulence in such shocks can spawn scattering
angles of the order of a degree or so.  It is this scenario that is the
one most probably pertinent to the 1--30 GeV leptons.

The inverse Compton gamma-ray signal measured by the {\it Fermi}-LAT is
probing leptons of energies in the TeV range. The spectroscopic demands
are now different: the 1--20 GeV index can be supplied by either
subluminal or superluminal shocks with SAS operating, provided that the
turbulence is not far from the Bohm limit.  LAS is also possible, but
would require highly superluminal conditions to effect the requisite
balance between large kinematic gains in shock-layer scatterings and
rapid convective losses downstream.  The X-ray spectrum samples the
highest energy electrons, in the super TeV range, that are subject to
strong cooling (burn-off) over the nebular lifetime.  Allowing for this
modification, the inferences for the injected lepton spectrum at the
PWTS are similar to those from the gamma-ray data.  A broadband picture
emerges that is highlighted in Atoyan \& Aharonian (1996): the electron
spectrum is convex (i.e. steepening) in the sense that \teq{\sigma} is
an increasing function of energy.  This is not difficult to accommodate
using the results from shock acceleration theory presented here, being
modeled by a modest transition from LAS at low energies to Bohm-domain
SAS at the highest pair energies.  The portions of the PWTS driving this
energization can be either superluminal or marginally subluminal.  It is
not hard to envisage turbulence that is slightly stronger at smaller
scales than larger ones that might precipitate this LAS \teq{\to} SAS
evolution with energy or Larmor radius in gyroresonant interactions. 
Yet in the near-term future, neither can observations resolve the
angular scales to demonstrate such, nor can plasma simulations probe the
wide dynamic ranges in lengthscales to validate such a scenario.

The simulation results presented here are for species of a single mass,
obviously applying to pure pair shocks.  It is natural to ask whether
they might differ if the abundance of ions is significant. The answer
must be deferred to future explorations of diffusive acceleration in
hydrogenic plasma shocks.  Yet it is expected that the the index results
should be the same unless the turbulence generation is different when
massive species are present.  At energies below 1 GeV, the gyrational
scales of protons and electrons of a given energy differ because the
protons are at most only mildly-relativistic.  This must lead to
significantly different gyroresonant interactions for \teq{e^-} and
\teq{p}. Furthermore, at these energies, charge separation cross shock
potentials are anticipated to play a profound role in energy exchange
between the two species (e.g. Baring \& Summerlin 2007). These two
contributions should provide substantial differences in injection
efficiency between pair shocks and hydrogenic or electron-ion ones. 
This injection issue is clearly salient for the overall prediction of
fluxes in different bands for PWNe.  However, at energies well above 10
GeV, the gyro-scale of a charge of a given energy is independent of its
mass, so that to leading order, turbulence generation and diffusion
characteristics should be similar in this domain for relativistic
\teq{e^{\pm}} and \teq{e-p} shocks.

Another question is whether or not the well-known non-linear spectral
concavity encountered in non-relativistic shocks that efficiently
accelerate charges (see Jones \& Ellison 1991; Ellison \& Double 2002;
Baring 2004, and references therein) might compete with and preclude the
spectral convexity that is demanded by the multiwavelength observations.
 Such non-linear enhancements of high energy particles arise for
distributions that have indices \teq{\sigma\lesssim 2}, where these
particles supply a sizeable portion of the total energy flux through the
shock, and thereby modify the global MHD shock structure.  While the
radio observations in PWNe access this domain, the distribution
convexity demanded by the gamma-ray and X-ray data must mute possible
non-linear modifications, so that they should play a more minor role
than in the non-relativistic shocks that illuminate supernova remnant
outer shells.

\section{Conclusions}
 \label{sec:conc}

This paper has outlined the key features of relativistic shock
acceleration that pertain to lepton injection at termination shocks into
pulsar wind nebulae. This shock is the most popular site for such
injection, because (i) all pairs emanating from the pulsar that travel
to the nebula must transit through this interface, (ii) it should be
turbulent and therefore an efficient injector/accelerator, and (iii) the
main characteristics of diffusive acceleration theory at shocks are
fairly well understood.  While it is quite possible that
pre-acceleration can arise in magnetic reconnection zones between the
pulsar light cylinder and the PWTS, such seed particles can be further
energized at the shock to the point of masking the signatures of
pre-acceleration. The historical models that developed the paradigm of
the PWTS as an injector predate refined studies of relativistic shock
acceleration over the last decade. As is evident here, these more recent
studies support such a paradigm in being able to generate the requisite
distribution indices to match the multiwavelength PWN observations
without appealing to unlikely situations concerning turbulence in the
shock layer.  The key issue that remains unresolved by theory is how
efficient injection arises from thermal energies in the PWTS.  Does it
occur for pure pair shocks, or is some baryonic loading necessary to
precipitate prolific  energization? Or, is a pre-acceleration seed
required to set the diffusive processes at the shock active all the way
to super-TeV energies?  Addressing such questions will require more
advanced simulations and theoretical analyses.  The answer will
illuminate the overall particle budget in pulsar wind nebulae, balancing
radiation luminosity, non-thermal particle energetics, and the wind
power from the pulsars that drives these fascinating systems.

\begin{acknowledgement}
This research was supported in part by National 
Science Foundation grant PHY07-58158 and NASA grant NNX10AC79G.  
\end{acknowledgement}

\end{document}